\documentclass[aps,twocolumn,prl,amsmath,amssymb,showpacs]{revtex4}
\usepackage{graphicx}
\usepackage{dcolumn}

\usepackage{bm}

\newcommand{\gma}{$\rm{Ga_{1-x}Mn_{x}As}$}

\newcommand{\tc}{$T_{\rm{C}}$}

\begin{document}
\title{Quasi-reversible Magnetoresistance in Exchange Spring Tunnel Junctions}
\author{M.\ Zhu}
\author{M.\ J.\ Wilson}
\author{P.\ Mitra}
\author{P.\ Schiffer}
\author{N.\ Samarth}\email{nsamarth@psu.edu}
\affiliation{Dept. of Physics and Materials Research Institute, The Pennsylvania State University, University Park PA 16802}

\begin{abstract}
We report a large, quasi-reversible tunnel magnetoresistance in exchange-biased ferromagnetic semiconductor tunnel junctions wherein a soft ferromagnetic semiconductor (\gma) is exchange coupled to a hard ferromagnetic metal (MnAs). Our observations are consistent with the formation of a region of inhomogeneous magnetization (an ``exchange spring") within the biased \gma~ layer. The distinctive tunneling anisotropic magnetoresistance of \gma~ produces a pronounced sensitivity of the magnetoresistance to the state of the exchange spring. 
\end{abstract}
\maketitle

The interplay between spin-polarized currents and ferromagnetic order is a fundamental problem that continues to attract much attention in contemporary condensed matter physics, with specific applications in spintronics. In this context, there is a particular need to identify model systems that could permit systematic experimental studies of spin-dependent transport through regions of non-uniform magnetization~\cite{Levy:1997lr,Foros_2008}. Here, we we describe spin-dependent tunneling studies of novel semiconductor-based magnetic tunnel junctions (MTJs) that contain an ``exchange spring'' produced by exchange coupling one of the ferromagnetic semiconductor layers with a magnetically harder (metallic) ferromagnet. In such exchange springs, the non-uniform magnetization of the softer ferromagnetic layer is twisted about an axis normal to the layer plane, akin to a partial domain wall \cite{Fullerton:1998:PRB}. We show that spin-dependent tunneling -- and thus the tunneling magnetoresistance (TMR) -- in these MTJ devices is very sensitive to the detailed magnetic configuration of the exchange spring because of the tunneling anisotropic magnetoresistance (TAMR) effect~\cite{Ruster:2005:PRL}. Although exchange spring configurations have been studied in metallic ferromagnetic heterostructures using techniques such as neutron scattering~\cite{odonovan:2002:PRL}, we are unaware of any measurements of spin-dependent tunneling in such systems. In addition to providing a new model system for studies of spin-dependent transport, the devices described here are a proof-of-concept demonstration of exchange-biased ferromagnetic semiconductor MTJs of relevance to semiconductor spintronics~\cite{Awschalom:2007:NatPhy}. 

The MTJ devices studied here are fabricated from heterostructure samples that are grown by low temperature molecular beam epitaxy on \textit{p}-type (001) GaAs substrates.  We first deposit a 150 nm thick Be-doped p-GaAs buffer layer (with \textit{p}$\sim$1$\times10^{19}$cm$^{-3}$). This is followed by a heterostructure comprised of two nominally identical \gma~ layers (with 30 nm thickness and $x \sim 6 \%$) separated by a tunnel barrier. In sample A, the tunnel barrier consists of 1 nm GaAs/ 4nm AlAs / 1nm GaAs spacer; the GaAs spacers prevent diffusion of Mn into the AlAs barrier. In samples B and C, the tunnel barriers consist of 4nm GaAs and 8 nm GaAs, respectively. The top \gma~ layer has an epitaxial overlayer of type-A MnAs (with thickness 10$\pm$2nm). The latter is a semimetallic ferromagnet that provides an exchange bias for the top \gma~ layer~\cite{Zhu:2007:APL}, making it harder to switch, while the bottom \gma~ layer is free to rotate in a small magnetic field. The function of the ferromagnetic MnAs layer is analogous to that of the antiferromagnetic layer commonly used in metallic MTJ devices as a pinning layer~\cite{Tsymbal:2003:TMR_review}.  Although it is possible to exchange bias \gma~ using an antiferromagnet, this has only been achieved using an insulator (MnO)~\cite{Eid:2004:APL}, thus precluding easy fabrication of perpendicular transport devices. 

We use photolithography, followed by a chlorine-based reactive ion etch, to fabricate cylindrical mesa MTJ devices, about 600 nm high and ranging in diameter from $10 \mu$m to 100$\mu$m. The TMR in these devices is measured using a pseudo-four-probe scheme, with separate voltage and current leads connected to top Ti/Au contacts and In contacts on the back of the substrate. The TMR is measured by applying a bias voltage and measuring the tunneling current using a Keithley 2410 source meter. The magnetic properties of the samples are measured using a Quantum Design superconducting quantum interference device (SQUID) magnetometer. In the magnetization and magnetoresistance measurements, the external magnetic field is applied in plane along $[1 1 \bar{2} 0]$, the easy axis of the type-A MnAs layer; this corresponds to the [110] direction of the GaAs substrate.

\begin{figure}[t]
\includegraphics[scale=1.2]{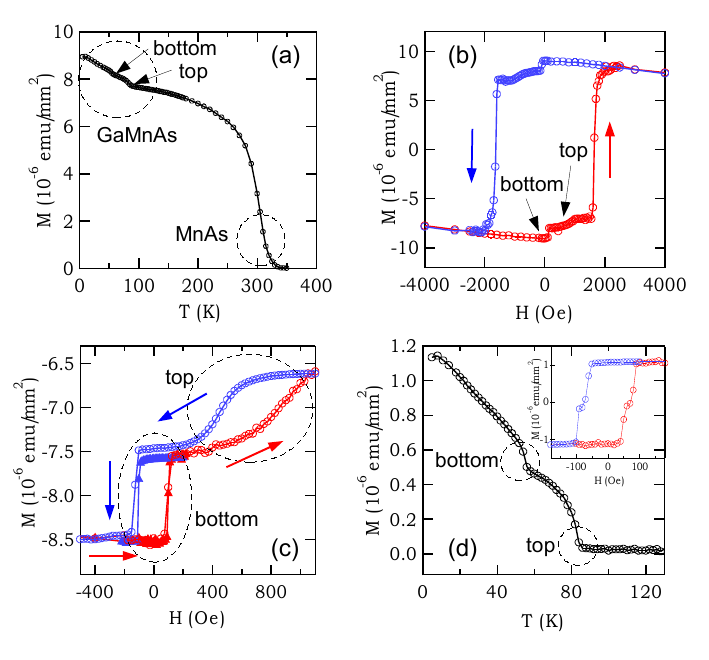}
\caption{(Color online) (a) Temperature-dependent magnetization $M(T)$ for sample C. (b) Major magnetization loop $M(H)$ for sample C, showing distinct switching fields for the bottom \gma~ layer, the top \gma~ layer and the MnAs layer ($\sim1.8$ kOe). (c) Minor magnetization loops for the same sample, measured after saturating the MnAs layer in a field of $- 20$ kOe. Solid triangles show the minor loop when only the bottom \gma~layer is switched. Hollow circles show the minor loop when both \gma~ layers switch. (d) $M(T)$ and $M(H)$(inset) of the same sample after chemical removal of the MnAs layer. All the hysteresis loops are measured at 4.2K.}
\end{figure}

Figure 1(a) shows the temperature-dependent remanent magnetization $M(T)$ for sample C. The other two samples show similar behavior. Although the two \gma~ layers have nominally identical compositions, they show different values of the Curie temperature (\tc). The $\sim 20$ K difference in \tc~ is readily seen in $M(T)$ when the top MnAs layer is chemically removed (Fig.1(d)). This enhancement in \tc~ for the upper \gma~ layer is attributed to the out-diffusion of hole-compensating Mn interstitial defects during the growth of MnAs. Figure 1(b) shows the major magnetization curve $M(H)$ in sample C, revealing different switching behaviors for the top and bottom \gma~ layers:  the bottom layer experiences a sharp transition while the magnetization of the top one gradually changes in external field due to the exchange coupling with the MnAs overlayer. 

Measurements of the minor $M(H)$ loops are shown in Fig.1(c).  We obtain these minor loops after first saturating the magnetization of MnAs layer in a -20 kOe field. Then, we sweep the magnetic field in a small field range to switch either the bottom \gma~ layer ($-300 \rm{Oe} \leq H \leq  300 \rm{Oe}$) or both layers ($-1 \rm{kOe} \leq H \leq 1 \rm{kOe}$). The bottom \gma~ layer shows a small coercivity with a square hysteresis loop that is symmetric around zero field. In contrast, the top \gma~ layer exhibits a large displacement ($\sim 600$ Oe) of the hysteresis loop due to the ferromagnetic exchange coupling to the MnAs layer~\cite{Zhu:2007:APL}. As a control experiment, we chemically etched away the MnAs layer on the same sample and measured $M(T)$ and $M(H)$; the inset to Fig. 1(d) shows that the coercivity of the top \gma~ layer is drastically reduced to $\sim 50$ Oe with a much sharper transition. 

\begin{figure}[t]
\includegraphics[scale=0.9]{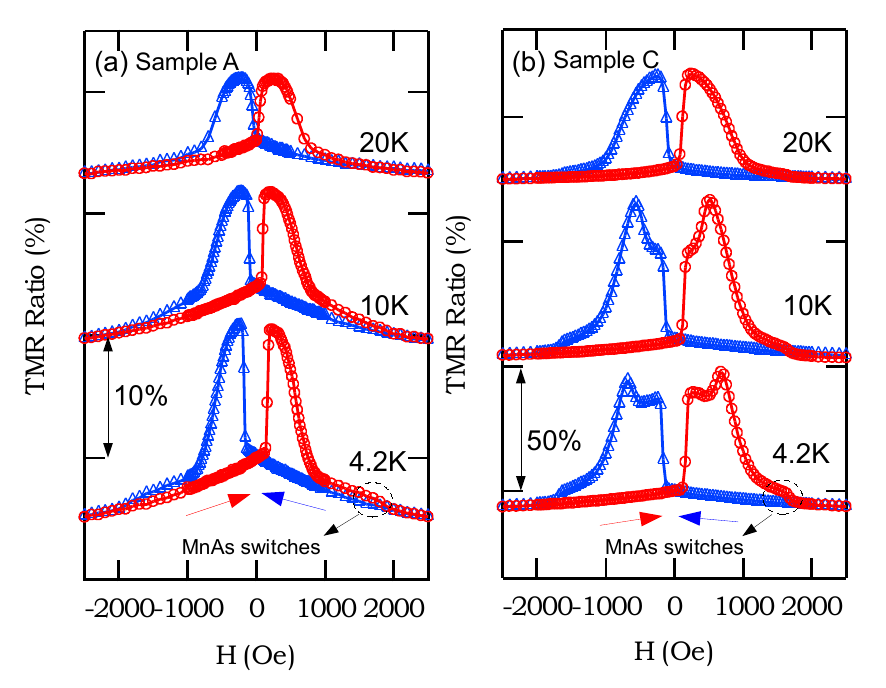}
\caption{(Color online) Major loop TMR at different temperatures in (a) a 100$\mu$m diameter mesa device patterned from sample A and (b) a 50$\mu$m device patterned from sample C. }
\end{figure}

Mesa devices fabricated from the three MTJ samples show distinct I-V characteristics (data not shown) due to different barrier heights and widths. Devices derived from sample A have the largest differential resistance (several k$\Omega$ for 100$\mu$m diameter mesas) and a strong non-linearity, as anticipated for an AlAs barrier with a height of $\sim$550 meV in the valence band~\cite{Tanaka:2001:PRL}. In contrast, since the height of the GaAs barrier is small ($\sim 100$meV ~\cite{Chiba:2004:PhysicaE}), devices made from sample B (4 nm wide barrier) display almost linear I-V characteristics, while those from sample C (8 nm wide spacer) show non-linear I-V behavior characteristic of a good tunnel barrier. In the rest of the paper, we focus on the TMR in two devices fabricated from sample A (100$\mu$m diameter) and sample C (50$\mu$um) where we have taken an extensive set of data. Measurements on other devices (of varying diameter) fabricated from these samples show qualitatively similar results. 

Figures 2(a) and (b) show the major loop TMR in samples A and C as a function of in-plane magnetic field over the temperatures range 4.2 K - 20 K. The magnetic field is applied along [110], parallel to the easy axis of the MnAs layer. The data shown in Fig. 2 are all taken at a  constant 4 mV dc bias; we find that the TMR ratio shows a monotonic decay with increasing bias (data not shown) and is symmetric in bias voltage polarity. The data show three distinct jumps in resistance, each corresponding to the independent switching of the three ferromagnetic layers. The sharp increase in resistance at low field corresponds to the switching of the lower \gma~ layer which creates an angle with respect to the upper \gma~ layer. The second jump in resistance occurs when the magnetization of the upper \gma~ layer switches direction and aligns with that of the lower \gma~ layer, creating a low resistance state. As we will discuss in more detail later, this process is more complex, resulting in a gradual decrease in TMR as the two layers come into alignment. For sample A, we find a peak TMR ratio as high as $\sim 12\%$ at 4.2 K, comparable to earlier reports in \gma~-based MTJs with thinner AlAs barriers~\cite{Tanaka:2001:PRL}. Sample C shows a much larger peak TMR ($\sim 40 \%$) at 4.2 K, consistent with more coherent tunneling through the GaAs barrier~\cite{Chiba:2004:PhysicaE}. We note that sample B (4 nm GaAs barrier) shows very weak TMR ($\sim 0.6 \%$ peak value, data not shown); we attribute this to the small tunneling barrier suggested by the linear I-V characteristics. In all three samples, the peak TMR decreases as expected with increasing temperature, due to the decreasing spin polarization of the \gma~ layers.

The third jump in resistance (indicated by arrows in Fig.2) occurs when the MnAs layer finally switches (at $\sim1800$ Oe), aligning its magnetization with that of the \gma~ layers. This additional lowering of the overall device resistance is accompanied by the closing of the major hysteresis loop. We observe this small decrease in resistance in devices derived from all three samples and attribute it to a spin valve effect due to spin dependent scattering at the MnAs/\gma~ interface~\cite{Zhu:2007:APL}. We note that once all the three ferromagnetic layers are aligned, the low resistance state is stable, and the TMR follows an irreversible path as the magnetic field is swept down. 

\begin{figure}[t]
\includegraphics[scale=1]{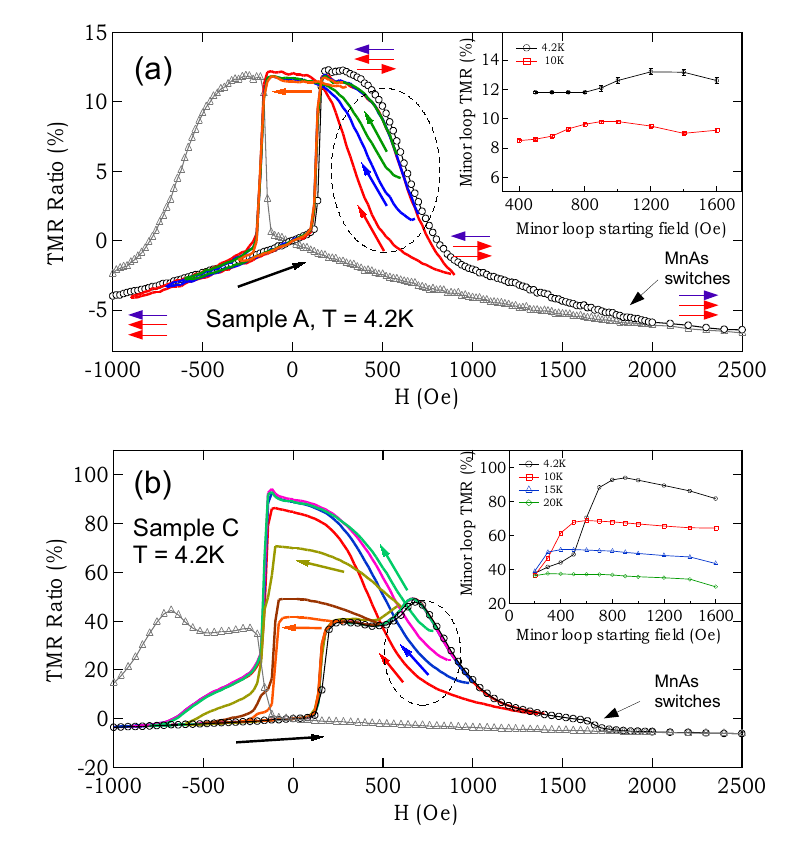}
\caption{(Color online)  Major loop TMR (circles for sweep up and triangles for sweep down) and minor loop TMR (lines) at 4.2 K for devices fabricated from (a) sample A and (b) sample C. The arrows indicate the direction of minor loops starting from different fields. The insets show the minor loop TMR ratio as a function of starting field at different temperatures for the two samples.}
\end{figure}

We now address the minor loop TMR shown as solid lines in Fig. 3. These sweeps are obtained by the following consistent protocol: we first saturate the magnetization of all three magnetic layers in a field of $- 3$ kOe; we then sweep to a starting field $+H_0$ and then sweep the field down in magnitude. The minor loop TMR has a complex behavior whose exact form is sensitive to the value of $H_0$. When the minor loop is initiated at values of $H_0$ just above the coercive field of the lower \gma~ layer, the TMR shows a standard irreversible minor hysteresis loop, with the high resistance state remaining stable as the field is swept down. However, if $H_0$ is in a range within which the top \gma~ layer has begun to switch direction (indicated within the dashed ellipses in Fig.3), the minor loop TMR is ``quasi-reversible." In sample A, the TMR retraces a path whose shape resembles that observed while the field is swept up, but with a field-offset whose value depends on $H_0$. In sample C, both the shape of the TMR and its amplitude are sensitive to $H_0$; most surprisingly, the peak value of the TMR increases dramatically during the down sweep, almost doubling in value compared with the peak TMR obtained during the sweep up. The behavior in both samples suggests that the magnetization of the upper \gma~ layer does not remain in a stable state while the magnetic field is swept down. We note that in the control sample where the MnAs layer is etched away, this quasi-reversible behavior disappears, showing a minor TMR similar to that reported by in earlier studies of \gma~ MTJs~\cite{Higo:2001:JAP}. We now develop a qualitative argument that links the minor loop TMR in these devices to the unwinding of an exchange spring in the upper \gma~ layer. 

\begin{figure}[b]
\includegraphics[scale=0.9]{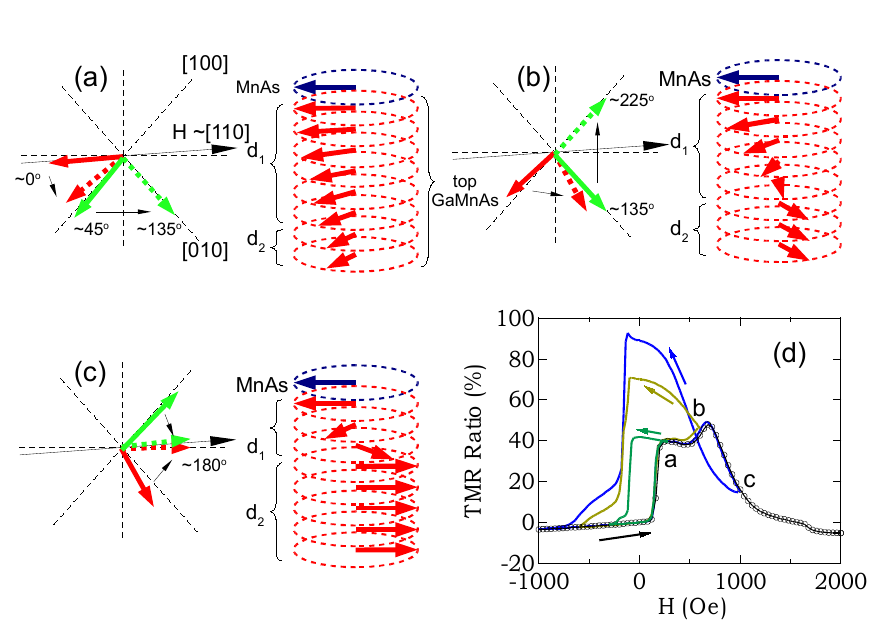}
\caption{(Color online) (a)(b)(c) Illustrations of the switching of magnetization in three different field ranges. The red (green) arrows represent the top (bottom) \gma~ layer respectively. The twisted spins in the top \gma~ layer are depicted in the right part of the figures, forming a spring-like configuration with variable depths. (d) Major loop (circles) and minor loops (lines) of sample C starting from different representing field regions (a,b,c)}
\end{figure}

We begin with a picture of the magnetization in the upper \gma~ layer. When the magnetic field is swept above a characteristic coercive field, we speculate that the exchange coupling of this layer with the MnAs layer results in an unusual exchange spring configuration that consists of a partial domain wall of width $d_{1}(H)$ in the vertical direction and a complete domain of width $d_{2}(H)$ produced by the anisotropy of the \gma~ layer (Fig. 4). It is reasonable to assume that both these domain widths ($d_1$ and $d_2$) are field-dependent. The width of the complete domain $d_{2}$ increases with magnetic field while the width of the partial domain wall $d_{1}$ diminishes. However, unlike complete exchange springs that exhibit fully reversible behavior with a disappearance of hysteresis \cite{Fullerton:1998:PRB}, the system studied here can follow a more complex quasi-reversible return path: when the magnetic field is swept down, the complete domain state may maintain a fixed width $d_2$ and rotate as a single domain, while the exchange spring unwinds. When combined with the inherent cubic and uniaxial magnetic anisotropy of \gma~, as well the the TAMR effect~\cite{Ruster:2005:PRL}, this results in minor loop TMR that can differ greatly in magnitude and shape from the major loop TMR.

This is best illustrated by the dramatic dependence of the minor loop peak TMR ratio on $H_0$ in sample C at the lowest temperatures. In Figs.4 (a)-(c), we illustrate the switching of magnetization for three different values of $H_0$; the respective TMR for these three minor loops is shown in Fig.4(d). The external field is applied along [110] with a small misalignment angle. Since the direction of the MnAs magnetization is fixed after saturating at $- 3$ kOe, the angles of the magnetization of \gma~ layers are described with respect to the MnAs magnetization. When the field is swept up after reversal, the first abrupt jump in TMR corresponds to the nucleation and propagation of a $\sim 90^{\circ}$ domain wall within the bottom \gma~ layer. This leads to an abrupt switching of its magnetization from $\sim 45^{\circ}$ to $\sim 135^{\circ}$, as indicated by the green arrows in Fig.4(a). The magnetization of the top layer also rotates inhomogeneously through a small angle, forming a relatively wide partial domain wall. When a minor loop starts from this configuration (sweeping down), the $M$ of the bottom layer simply switches back to $\sim 45^{\circ}$, giving a square hysteresis loop (green curve in Fig.4(d)). In this regime, the TMR magnitude does not vary much with the reversible unwinding of the slightly twisted exchange spring in the upper layer.

For larger values of $H_0$ that lie in the major loop TMR plateau ($200$ Oe $\lesssim H_0 \lesssim 700$ Oe), the magnetization of the bottom layer gradually rotates to the second energy minimum, aligning itself at an angle of $\sim 225^{\circ}$ (Fig.4(b)). The presence of a uniaxial anisotropy in \gma~ (in addition to the cubic anisotropy) causes this second $90^{\circ}$ switching through coherent rotation, rather than domain wall nucleation. This is accompanied by an additional twisting of the exchange spring in the upper layer. We speculate that the complex plateau in the major loop over that field range is a result of the interplay of both top and bottom \gma~ layers rotating in the same direction. When the minor loop is initiated with $H_0$ in this regime, the magnetization of the lower layer is pinned in its energy minimum at $\sim 225^{\circ}$ until the field reverses. In the upper layer, however, the complete domain rotates coherently to an energy minimum at $\sim 45^{\circ}$. Because of the TAMR effect and the larger angle between the two magnetizations, it yields a larger peak TMR (yellow and blue curves in Fig.4(d)) than that seen during the major loop. Further, the width $d_2(H)$ of the complete domain -- and correspondingly the magnetization along $[\bar100]$ -- increases with $H_0$. Using the heuristic Julliere model~\cite{Julliere:1975:PL}, this increase in magnetization (and the accompanying increase in spin polarization) explains the increase in the TMR peak value with increasing $H_0$ (inset in Fig.3(b)).  The increase in peak TMR with $H_0$ is however not monotonic and reaches a maximum value for $H_0 \sim 700$ Oe, beyond which we observe a slight decrease in the peak TMR (inset in Fig.3(b)). We may reasonably assume that even though the width $d_2$ of the complete domain increases with $H_0$, only a portion of this complete domain can rotate coherently, limited to a critical thickness $d_{c}$. Consequently, the peak TMR is maximum when $d_2 = d_c$. We do not have a detailed explanation for the slight decrease in peak TMR at higher values of $H_0$, but it is consistent with the existence of pinning sites for small domains.  

Our qualitative model also explains the observed temperature dependence of the minor loop TMR (see inset to Fig.3(b)). With increasing temperature, the peak value of the minor loop TMR of sample C becomes less dependent on the starting field, similar to the behavior observed at all temperatures in sample A (inset to Fig.3(a)). We note that -- at these higher temperatures -- the overall character of the minor loop TMR in sample C still remains quasi-reversible (data not shown) and resembles that obtained for sample A at all temperatures. We attribute these observations to a rapid decrease in the cubic anisotropy with increasing temperature relative to the change in the uniaxial anisotropy~\cite{Sawicki:2005:PRB}: this reduces TAMR contributions by removing the lack of symmetry between the two easy axes. A puzzling aspect of our data is the contrast in minor loop TMR between samples A and C at the lowest temperature, particularly given that the only structural difference between the two samples is the tunnel barrier. We do not currently have a convincing explanation for this difference, but note that the magnetic anisotropy of \gma~ is a sensitive function of hole density. Studies of epitaxially grown heterostructures containing \gma~ have shown a large variability in hole-compensating Mn interstitial concentrations that depends on the details of the heterostructure. It is thus possible that the lack of TAMR in sample A may arise from a lack of cubic anisotropy in the lower \gma~ layer, resulting in Fermi-level driven differences in Mn interstitial diffusion during the overgrowth of AlAs. It is also possible that contributions of TAMR are sensitive to the nature of the tunnel barrier, although calculations of TAMR for GaAs and AlAs tunnel barriers do not suggest an obvious explanation\cite{Sankowski:2007:PRB}.

This research has been supported by the ONR MURI program N0014-06-1-0428 through subcontract KK6135 with the University of California, Santa Barbara. This work was performed in part at the Penn State Nanofabrication Facility, a member of the NSF National Nanofabrication Infrastructure Network. 


\end{document}